\documentclass[11pt,a4paper]{article}
\usepackage{epsfig,amsmath,amsfonts}%
\usepackage{ifthen}

\def \beq {\begin{equation}}
\def \eeq {\end{equation}}
\def \beqar {\begin{eqnarray}}
\def \eeqar {\end{eqnarray}}
\def \beqa* {\begin{eqnarray*}}
\def \eeqa* {\end{eqnarray*}}
\def \pa {\partial}

\def \d {\,{\rm{d}}}
\def \polytrop {\nu}

\def\refart#1#2#3#4#5#6#7{{#1}, \textit{#2}, \textit{#3} \textbf{#4} {(#6)} {#5}{\ifthenelse{\equal{#7}{}}{}{ [#7]}}}%

\def\refproc#1#2#3#4#5#6{{#1}, \textit{#2}, in \textit{#3}, #4 {(#6)}, pg.~{#5}}%

\title{Statefinder analysis of the superfluid Chaplygin gas model}
\author{V.\,A.~Popov}%

\begin{document}

\maketitle

\begin{center}\small \textit{Department of General Relativity and
Gravitation,\\Kazan Federal University,\\Kremlyovskaya st. 18,
Kazan 420008, Russia}\end{center}

\begin{center}\small Email address: vladipopov@mail.ru\\
\end{center}

\begin{abstract}
The statefinder indices are employed to test the superfluid
Chaplygin gas (SCG) model describing the dark sector of the
universe. The model involves Bose-Einstein condensate (BEC) as
dark energy (DE) and an excited state above it as dark matter
(DM). The condensate is assumed to have a negative pressure and is
embodied as an exotic fluid with the Chaplygin equation of state.
Excitations forms the normal component of superfluid. The
statefinder diagrams show the discrimination between the SCG
scenario and other models with the Chaplygin gas and indicates a
pronounced effect of the DM equation of state and an indirect
interaction between their two components on statefinder
trajectories and a current statefinder location.%
\end{abstract}

\noindent{\small \emph{Keywords:} accelerated expansion, Dark
Energy, Dark Matter, relativistic superfluid, Chaplygin gas,
statefinder

}

\newpage

\section{Introduction}

The energy content of the Universe is a fundamental issue in
cosmology. Observational data, such as Type Ia Supernovae
(SNIa)~\cite{Riess&Perlmutter}, Cosmic Microwave Background
(CMB)~\cite{BOOMERanG&WMAP} and Large Scale Structure~\cite{SDSS},
are evidence of accelerating flat Friedmann-Robertson-Walker
model, constituted of about 1/3 of baryonic and dark matter  and
about 2/3 of a dark energy component.

The essential feature of DE is that its pressure must be negative
to reproduce the present accelerated cosmic expansion. There are a
few candidates for DE incorporated in competing cosmological
scenarios. The simplest DE model, the cosmological constant, is
indeed the vacuum energy with the equation of state $p=-\rho$. A
number of models, such as quintessence \cite{Wetterich1},
k-essence \cite{Armendariz1}, phantom \cite{Caldwell1} and etc.,
are based on scalar field theories. Braneworld models explain the
acceleration through the five-dimensional general
relativity~\cite{braneworld}. The Chaplygin gas model, also
denoted as quartessence, exploits a negative pressure fluid, which
is inversely proportional to the energy density
\cite{Kamenshchik1}. For a more detail review of DE models and
references see \cite{Copeland1}. Besides, there are models
unifying DE and DM, including a some kind of scalar fields
\cite{UniScFlds}, generalized Chaplygin gas (GCG)
\cite{Bento&Bilic,Bento2} and superfluid Chaplygin gas (SCG)
\cite{SCG 1}.

In order to differentiate these various DE models, Sahni et al.
\cite{Sahni-sf-intro} introduce a new geometrical diagnostic pair
$\{r, s\}$, called \emph{statefinder}, which involves the third
order derivative of the scale factor with respect to time. Its
important attribute is that the spatially flat $\Lambda$CDM has a
fixed point $\{r, s\}=\{1,0\}$. Departure of a DE model from this
fixed point is a good way of establishing the `distance' of this
model from flat $\Lambda$CDM. The statefinder diagnostic has been
also applied to several DE models
\cite{Alam-sf,sf-phantom,Kam-sf-in-Chap,sf-holo,sf-interact,sf-etc}
to differentiate them from $\Lambda$CDM and one from other. In
addition the values of the statefinder pair can be extracted from
data of SuperNova Acceleration Probe (SNAP) type experiments
\cite{Sahni-sf-intro,Alam-sf} to obtain constraints on the models.

In this letter the statefinder diagnostic is applied to the SCG
model developed in \cite{SCG 1}. It represents the dark sector of
the universe as a superfluid where the superfluid condensate is
considered as DE and the normal component is interpreted as DM.
The model is based on the action
\beq\label{Action}
S=\int \left(-\frac{R}{16\pi G}+{\cal L}\right)\sqrt{-g}\d^4x\, ,
\eeq
where
Lagrangian ${\cal L}$ %
associated with a generalized hydrodynamic pressure function
depends only on one variable if we consider pure condensate, and
on three variables when we include the excitation gas. To provide
the accelerated expansion the negative pressure of the superfluid
background obeys Chaplygin's equation of state. In Sec.~\ref{Sec:
SF dynamics} and \ref{Sec: Universe with BEC} the SCG model is
briefly outlined. The statefinder evolution and differentiation
between SCG model and another models with the Chaplygin gas are
discussed in Sec.~\ref{Sec: Statefinder}. The metric signature
$(+---)$ is adopted in this work.

\section{Relativistic superfluid dynamics}
\label{Sec: SF dynamics}

An efficient approach to description of the excited state is
two-fluid hydrodynamics. This theory does not depend on details of
microscopic structure of the quantum liquid and exploits effective
macroscopic quantities. In the theory there exist two independent
flows, the coherent motion of the ground state named a superfluid
component, and a normal component produced  by the quasiparticle
gas. For this reason  it is necessary to increase the number of
independent variables in the generalized pressure
(\ref{PressureInCondensate}) from one to three
\cite{Khalatnikov&Carter}. They correspond to three scalar
invariants which can be constructed from the pair of independent
vectors, namely superfluid $\mu_\alpha$ and thermal
$\theta_\alpha$ momentum covectors so that the general variation
of the generalized pressure in a fixed background is
\beq\label{dP1}
\delta P= \delta {\cal L} = n^\alpha \delta \mu_\alpha + s^\alpha
\delta \theta_\alpha\, .
\eeq
The coefficients $n^\alpha$ and $s^\alpha$ are to be interpreted
as particle number and entropy currents correspondingly. By virtue
of its invariance the pressure is given as a function of three
independent variables,  $I_1=\frac{1}{2}\mu_\alpha
\mu^\alpha,\,I_2=\mu_\alpha
\theta^\alpha,\,I_3=\frac{1}{2}\theta_\alpha \theta^\alpha$.
Taking the derivatives of the pressure, one finds
\beq\label{lin}
n^\alpha = \frac{\pa P}{\pa I_1} \mu^\alpha + \frac{\pa P}{\pa I_2} \theta^\alpha, \qquad%
s^\alpha = \frac{\pa P}{\pa I_2} \mu^\alpha + \frac{\pa P}{\pa
I_3} \theta^\alpha.
\eeq

As soon as the generalized pressure is the Lagrangian density in
the action (\ref{Action}) its variation with respect to the metric
gives the energy-momentum tensor
\beq
T_{\alpha\beta}=\frac{\pa P}{\pa I_1}\mu_\alpha \mu_\beta +
\frac{\pa P}{\pa I_2}(\mu_\alpha \theta_\beta + \theta_\alpha
\mu_\beta) +\frac{\pa P}{\pa I_3}\theta_\alpha \theta_\beta -
Pg_{\alpha\beta}\, .
\eeq

Instead of the thermal momentum $\theta_\alpha$ let us introduce
an inverse temperature vector
$\beta^\alpha=s^\alpha/(s^\beta\theta_\beta)$ which one uses as
the independent vector together with the superfluid momentum
$\mu_\alpha$ since they are comoving to the excitation gas and the
condensate respectively. Corresponding unit 4-velocities are
\beq\label{Velocities}
U^\alpha=\frac{\beta^\alpha}{\sqrt{\beta^\beta \beta_\beta}}\, ,
\qquad V^\alpha=\frac{\mu^\alpha}{\sqrt{\mu^\beta \mu_\beta}}\, .
\eeq

In place of the scalars $I_1,\ I_2,\ I_3$ one uses new three
invariants, a chemical potential $\mu=\sqrt{\mu^\beta \mu_\beta}$,
scalar $\gamma=V_\alpha U^\alpha$ associated with the relative
motion of the components, and inverse temperature with respect to
the reference frame comoving to the normal component
$\beta=\sqrt{\beta^\beta \beta_\beta}\,$.

Using (\ref{lin}) and (\ref{Velocities}) the energy-momentum
tensor and the particle number current are readily represented as
\beqar
n^\alpha &=& n_{\rm c}V^\alpha + n_{\rm n}U^\alpha ,\label{sfPN}\\
T_{\alpha\beta} &=& \mu n_{\rm c} V_\alpha V_\beta + W_{\rm n}
U_\alpha U_\beta - P g_{\alpha\beta}\, .\label{sfEMT}
\eeqar

The ground state is described by the generalized hydrodynamic
pressure function depending only on $\mu$. We will consider the
condensate with the function of the generalized pressure in the
form
\beq\label{PressureInCondensate}
P(\mu)=p_{\rm c}=-\sqrt{A-\lambda\mu^2}\, .
\eeq
It leads to the following particle and energy densities
\beq
n_{\rm c}=\frac{\lambda\mu}{\sqrt{A-\lambda\mu^2}}\, ,\qquad%
\rho_{\rm c}=\frac{A}{\sqrt{A-\lambda\mu^2}}\, .
\eeq
It is easy to see that if to eliminate the chemical potential
$\mu$ one can obtain
\beq\label{ChaplyginEOS}
n_{\rm c}=\sqrt{\frac{\lambda}{A}}\sqrt{\rho_{\rm c}^2-A}\, ,\qquad%
p_{\rm c}=-\frac{A}{\rho_{\rm c}}\, ,\qquad%
\eeq
and the adiabatic speed of sound
\beq\label{SoundSpeed}
c_{\rm s}^2=\frac{\d p_{\rm c}}{\d\rho_{\rm c}}=\frac{A}{\rho_{\rm
c}^2}\, .
\eeq
The equation of state (\ref{ChaplyginEOS}) is uniquely proper to
the Chaplygin gas suggested by Kamenshchik et
al.~\cite{Kamenshchik1} as an alternative to quintessence and
developed by a number of authors for description of the dark
sector of the universe \cite{Bento&Bilic}. In contrast to these
works where pressure of the Chaplygin gas is formed by both DE and
DM, this model implies that the equation of state
(\ref{ChaplyginEOS}) concerns with only BEC which is interpreted
as DE. Note that the generalized pressure
(\ref{PressureInCondensate}) can be obtained from the Lagrangian
\beq\label{LagrangianBasic}
{\cal L} = \pa_\nu \phi^* \pa^\nu \phi -M \left( \frac{\phi^*\phi}{\lambda} +
\frac{\lambda}{\phi^*\phi} \right)
\eeq
for a complex scalar field $\phi$ in the WKB-approximation \cite{SCG 1}.

The interesting aspect of the pressure function (\ref{PressureInCondensate}) is that it is a hydrodynamical representation of the generalized Born-Infeld Lagrangian
\beq
{\cal L}_\text{BI} = -\sqrt A \sqrt{1-\pa_\nu \theta \, \pa^\nu \theta}
\eeq
describing a (3+1)-dimensional brane universe with the scalar field $\theta$ in a (4+1)-dimensional bulk \cite{Bento&Bilic}.

More detail information about the excited state can be derived
from statistical description of the elementary excitations. The
quasiparticle energy spectrum has a significant nonlinear
dispersion at high energy, and therefore completely relativistic
description has been carried out only for a low energy
excitations, phonons \cite{Carter2,Popov1}. Based on the
relativistic kinetic theory of the phonon gas \cite{Popov1} one in
particular can obtain
\beq\label{EOSnn}
\frac{\mu n_{\rm n}}{\gamma}=(1-c_\text{s}^2)W_{\rm n}
\eeq
when phonons prevail over another sorts of quasiparticles.

Let us assume that the generalized pressure function is separated
as follows:
\beq\label{AnsatzSeparatedCondensate}
P(\mu,\beta,\gamma)=p_{\textrm{c}}(\mu)+p_{\textrm{n}}(\mu,\beta,\gamma)
\eeq
and
\beq\label{PnPhonon}
p_{\textrm{n}}(\mu,\beta,\gamma)=\frac{B(\mu)}{\beta^{\polytrop+1}
\left( 1- \gamma^2 \left(1-c_\text{s}^2\right)
\right)^{(\polytrop+1)/2}}
\eeq
that is inspired by the equilibrium pressure for the phonon gas
\cite{Carter2} corresponding to $\polytrop=3$. Eq.~(\ref{PnPhonon})
leads to the barotropic equation of state
\beq\label{AnsatzPn}
p_{\rm n}=\frac{1- \gamma^2
\left(1-c_\text{s}^2\right)}{\polytrop+1}W_{\rm n}\, .
\eeq


Moreover, the ansatz~(\ref{AnsatzSeparatedCondensate}) and (\ref{PnPhonon}) proves to be convenient for a number of reasons. The equation of state (\ref{PnPhonon}) makes possible to avoid a detail consideration of the full quasiparticle spectrum. Manipulating the sole parameter $\polytrop$ one can simulate a behavior of the normal component as DM. It also simplifies the following study of the cosmic evolution. This ansatz keeps the condensate self-dependent, i.e. eqs.~(\ref{ChaplyginEOS}) and (\ref{SoundSpeed}) remain  valid for the condensate in the framework of two-fluid dynamics and allow to naturally divide the total energy density into DE and DM fractions.

We restrict our consideration to the equation of state
(\ref{AnsatzPn}) for the normal component situated between the
dust one and the stiff one. It is evident from
eq.~(\ref{AnsatzPn}) that this constraint implies $\polytrop\ge
1$.

\section{Universe with SCG}
\label{Sec: Universe with BEC}

The cosmic medium is now regarded as a matter which particularly
is in the BEC state and its particle number current and
energy-momentum tensor have the form (\ref{sfPN}) and
(\ref{sfEMT}) where the superfluid background obeys the equation
of state (\ref{ChaplyginEOS}) and the excited state is described
by the relations (\ref{EOSnn}) and (\ref{AnsatzPn}).

Let us consider a homogeneous and isotropic spatially flat
universe. In this case the superfluid and normal velocities are
equal and thus $\gamma=1$. Einstein's equations then reduce to
\beq\label{EinsteinEqs}
3\frac{\dot a^2}{a^2} = 8\pi G \rho_{\textrm{tot}}\, , \qquad
-6\frac{\ddot a}{a} = 8\pi
G(3p_{\textrm{tot}}+\rho_{\textrm{tot}})\, ,
\eeq
where $\rho_{\textrm{tot}}$ consists of the condensate density
$\rho_{\textrm{c}}$ and the normal one
$\rho_{\textrm{n}}=W_{\textrm{n}}-p_{\textrm{n}}$ that are
interpretable as DE and DM densities respectively, and
$p_{\textrm{tot}}=p_{\textrm{c}}+p_{\textrm{n}}=P$. In accordance
with the integrability conditions of the Einstein equations we
require local energy-momentum conservation $\nabla_\mu
T^{\mu\nu}=0$ that yields
\beq\label{EnergyConservation}
\dot\rho_{\textrm{tot}} + 3\frac{\dot
a}{a}(p_{\textrm{tot}}+\rho_\textrm{tot})=0\, .
\eeq
The interaction between DE and DM is implicitly included in
equation (\ref{EnergyConservation}) and also in particle number
conservation $\nabla_\mu n^\mu = 0$ that leads to
\beq\label{ParticleConservation}
\dot n_{\textrm{tot}} + 3\frac{\dot a}{a}n_{\textrm{tot}}=0
\quad\Longrightarrow\quad%
n_\textrm{c}+n_\textrm{n} = \frac{n_0}{a^3}\, , \
n_0=\mbox{const}.
\eeq

Taking into account the expressions (\ref{EOSnn}),
(\ref{AnsatzPn}) with $\gamma=1$ and (\ref{ParticleConservation}),
eqs.~(\ref{EinsteinEqs}) and (\ref{EnergyConservation}) are
reduced to following two dimensionless equations:
\beqar
&&%
 3(1+\polytrop)\frac{\dot a^2}{a^2} = \frac{1}{\rho_\text{c}}+
\frac{k}{a^3} \left(
\frac{\polytrop\rho_\text{c}}{\sqrt{\rho_\text{c}^2-1}} +
\frac{\sqrt{\rho_\text{c}^2-1}}{\rho_\text{c}} \right),
\label{Eq1}\\&&%
3\frac{\dot a}{a}\left(1+\polytrop-
            \frac{k}{a^3}\frac{1}{\sqrt{\rho_\text{c}^2-1}}\right) +
            \frac{\dot\rho_\text{c}}{\rho_\text{c}}
            \left(1- \frac{k}{a^3}\left(\frac{1}{\sqrt{\rho_\text{c}^2-1}}-
            \frac{\polytrop\rho_\text{c}^2}{(\rho_\text{c}^2-1)^{3/2}}\right) \right)=0\, ,
\label{Eq2}
\eeqar
where the notation $\rho_\text{c}$ is used now for the
dimensionless energy density $\rho_\text{c}/\sqrt A$ as well as
$\rho_\text{n}$ will be used for $\rho_\text{n}/\sqrt A$ and etc.
The dimensionless time variable $t'$ is connected with real time
$t$ as $t'=\sqrt{8\pi G A^{1/2}}\,t$ and $k=n_0/\sqrt{\lambda}$.

In the formal limit $\polytrop\to\infty$ eqs.~(\ref{Eq1}) and
(\ref{Eq2}) are solved analytically. As obvious from
(\ref{AnsatzPn}) the quasiparticle pressure is neglected and DM
behaves as dust-like matter. In this case eq.~(\ref{Eq2}) yields
the condensate energy density in the form
\beq\label{rho_c:dust}
\rho_{\rm c} = \sqrt{\frac{k^2}{(a^3+\kappa_0)^2}+1}\, ,
\eeq
and the DM energy density is governed by the law
\beq\label{rho_n:dust}
\rho_{\textrm{n}} = \frac{\kappa_0}{a^3}
\sqrt{\frac{k^2}{(a^3+\kappa_0)^2}+1}\,.
\eeq
It is clear from (\ref{rho_c:dust}) and (\ref{rho_n:dust}) the
integration constant $\kappa_0$ is the current ratio between the
DM and DE energy densities.

At the beginning stage (i.e. for small $a$) the total energy
density is approximated by $\rho_{\textrm{tot}}\propto a^{-3}$
that corresponds to a universe dominated by dust-like matter. The
same behavior is a feature of  the Chaplygin gas
\cite{Kamenshchik1} but even though in this model the condensate
has the same equation of state, such  dependence is due to the
normal component.

At the late stage (i.e. for large $a$) $\rho_{\textrm{tot}}\to 1$.
Separating now DE and DM contributions one finds the subleading
terms are
\beq
\rho_{\rm c}  \sim  1+\frac{k^2}{2}a^{-6}, \qquad
\rho_{\textrm{n}}  \sim  \frac{\kappa_0}{a^3}\, ,
\eeq
whereas the scale factor time evolution corresponds to de Sitter
spacetime, namely, $a\propto e^{t'/\sqrt{3}}$.

\begin{figure}[t]
\begin{center}%
\includegraphics[width=.5\textwidth]{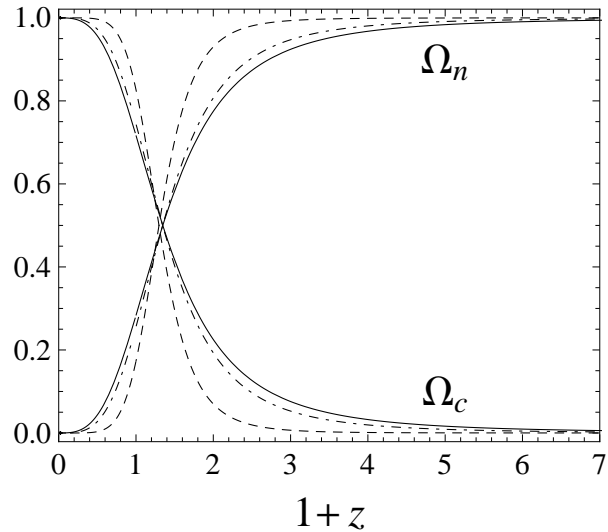}%
\end{center}%
\caption{The ratio of the energy density to the critical density
for both components of SCG as a function of the redshift $z$ for
$k=.2$ and the current value of $\rho_\text{c}(t_0)=1.01$. The
quantity $\polytrop$ varies as 1 (dot-dashed line), 5 (dashed
line) and 25 (solid line).}\label{FigOmega}
\end{figure}

When $\polytrop$ has a finite value the asymptotic behavior of
$\rho_\text{c}$ is the same as it is found in the case with pressureless DM  while
$\rho_\text{n}\propto a^{-3(1-1/\polytrop)}$ for small $a$, and
$\rho_\text{n}\propto a^{-3(1+1/\polytrop)}$ for large $a$. As
before the universe falls within de~Sitter phase in the far
future. At the intermediate stage eqs.~(\ref{Eq1}) and (\ref{Eq2})
are solved numerically. The figure~\ref{FigOmega} depicts an
evolution of the normalized energy densities $\Omega_{\rm c}$ and
$\Omega_{\rm n}$ of DE and DM respectively. The curves are plotted
for different values of $\polytrop$ and fixed $k$ and the current
value of $\rho_\text{c}$. The latter is close to 1 to provide the
correspondence with the current observational value of the DE
fraction $\Omega_{\rm c}\approx 0.7$.

Photometric observations of apparent Type Ia supernovae attests
that the recent cosmological acceleration commenced at
$0.3<z_\text{T}<1$ \cite{Wang1}. To satisfy this condition the
quantity $\polytrop$ have to be large, $\polytrop \geq 20$. This
implies a lower effective  sound speed\footnote{It is known as a
second sound speed in the superfluid theory.} for the normal
component evolving as $c_\text{s}/\sqrt\polytrop$. In this case
properties of the normal component are close to CDM. Note, that in
superfluid helium a lower second sound speed is provided by
quasiparticles from the nonlinear part of the energy spectrum
(such as rotons). To develop more realistic model, a wide
quasiparticle spectrum should be taken into account. In the
context of the pure phonon consideration they are not taken into
account and their influence is simulated with a large value of
$\polytrop$.

\section{Statefinder diagnostic}
\label{Sec: Statefinder}

In this section we focus our attention to the statefinder
diagnostic of the SCG model. The parameter pair $\{ r,s \}$ called
''statefinder'' was introduced  by Sahni et al.
\cite{Sahni-sf-intro} for the purpose to differentiate between
competing cosmological scenarios involving DE. The statefinder
test is a geometrical one based on the expansion of the scale
factor $a(t)$ near the present time $t_0$:
\beq
a(t)=1 + H_0(t-t_0) - \frac{1}{2}q_0H_0^2(t-t_0)^2 +
\frac{1}{6}r_0H_0^3(t-t_0)^3 + \ldots \, ,
\eeq
where $a(t_0)=1$ and $H_0,\ q_0,\ r_0$ are the current values of
the Habble constant $H=\dot a/a$, deceleration factor $q=-\ddot
a/aH^2$ and the former statefinder index $r=\dddot a/aH^3$
respectively. The latter index $s$ is the combination of $r$ and
$q$: $s=(r-1)/3(q-1/2)$.

Since the different cosmological models exhibit qualitatively
different trajectories in the $r-s$, $q-r$ or $q-s$ planes, the
statefinder diagnostic is a good tool to distinguish them. The
remarkable property of the pair $\{ r,s \}$ is that the
$\Lambda$CDM corresponds to the fixed point $\{ r,s \}=\{ 1,0 \}$.

In fact, the statefinder diagnostic has been successfully used to
test a number of models such as the cosmological constant, the
quintessence \cite{Alam-sf}, the phantom \cite{sf-phantom}, the
Chaplygin gas \cite{Kam-sf-in-Chap,Alam-sf}, the holographic dark
energy models \cite{sf-holo}, the interacting dark energy models
\cite{sf-interact} and etc. \cite{sf-etc}. On the other hand the
statefinder indices can be estimated from SNAP type experiment
\cite{Sahni-sf-intro,Alam-sf} to examine DE models from the
observational data.  In what follows we will calculate the
statefinder parameters for the SCG model and plot the evolution
trajectories in the statefinder planes.

The deceleration factor and the statefinder pair can also be
expressed as
\beqar
q &=& \frac{1}{2}\left( 1+3\frac{p_\text{tot}}{\rho_\text{tot}}
\right), \label{qPress}\\
r &=& 1 + \frac{9(\rho_\text{tot}+p_\text{tot})}{2\rho_\text{tot}}
\, \frac{\dot p_\text{tot}}{\dot\rho_\text{tot}} \, , \label{rPress}\\
s &=& \frac{(\rho_\text{tot}+p_\text{tot})}{p_\text{tot}} \,
\frac{\dot p_\text{tot}}{\dot\rho_\text{tot}} \, ,\label{sPress}
\eeqar
where the overdot denotes the the derivative with respect to the
time. Using the dimensionless energy density and pressure
expressions (\ref{rPress}) and (\ref{sPress}) can rewritten as
\beqar
r &=&  1 -
\frac{3\sqrt 3}{2} \, \frac{\dot p_\text{tot}}{\rho_\text{tot}^{3/2}} \, , \label{rSCGgen}\\
s &=&  - \frac{\dot p_\text{tot}}{p_\text{tot}
\sqrt{3\rho_\text{tot}} } \, ,\label{sSCGgen}
\eeqar
where the overdot denotes now the the derivative with respect to
the dimensionless time.

It is because the SCG model uses the unified conservation laws for
DM and DE and the DM pressure is non-zero in general, that the
expressions (\ref{rSCGgen}) and (\ref{sSCGgen}) are best suited to
calculate the statefinder indices to analyze the impact of the SCG
parameters on the statefinder location and reveal the difference
in the statefinder evolution for some models with the Chaplygin
gas.

First we consider the special case $\polytrop\to\infty$ when the normal
component is pressureless and the DE and DM energy densities
evolve according to the expressions (\ref{rho_c:dust}) and
(\ref{rho_n:dust}) respectively. In this case one can obtain
explicit dependance $r(s)$ but in view of it complexity it is
expressed  as follows
\beqar
q &=& \frac{1}{2} \left( 1 - \frac{3 a^3 \left( a^3+\kappa_0
\right) }{k^2
+ \left( a^3+ \kappa_0 \right)^2 } \right),\\%
r &=& 1 + \frac{9}{2}\,\frac{k^2 a^6}{\left( k^2 + \left( a^3+
\kappa_0
\right)^2 \right)^2}\, ,\\%
s &=& -\frac{k^2 a^3}{\left( a^3+ \kappa_0 \right)^2 \left( k^2 +
\left( a^3+ \kappa_0 \right)^2 \right)}\, ,
\eeqar
where the scale factor $a$ appears as a natural parameter.

The value of $k$ is directly related to the scale factor
corresponding to the transition from the deceleration to the
acceleration. It is agreed that the transition occurs at
$0.3<z_\text{T}<1$ \cite{Wang1} resulting in the restriction
$k<0.652$ for dust-like normal component.

\begin{figure}[t]
\includegraphics[width=.45\textwidth]{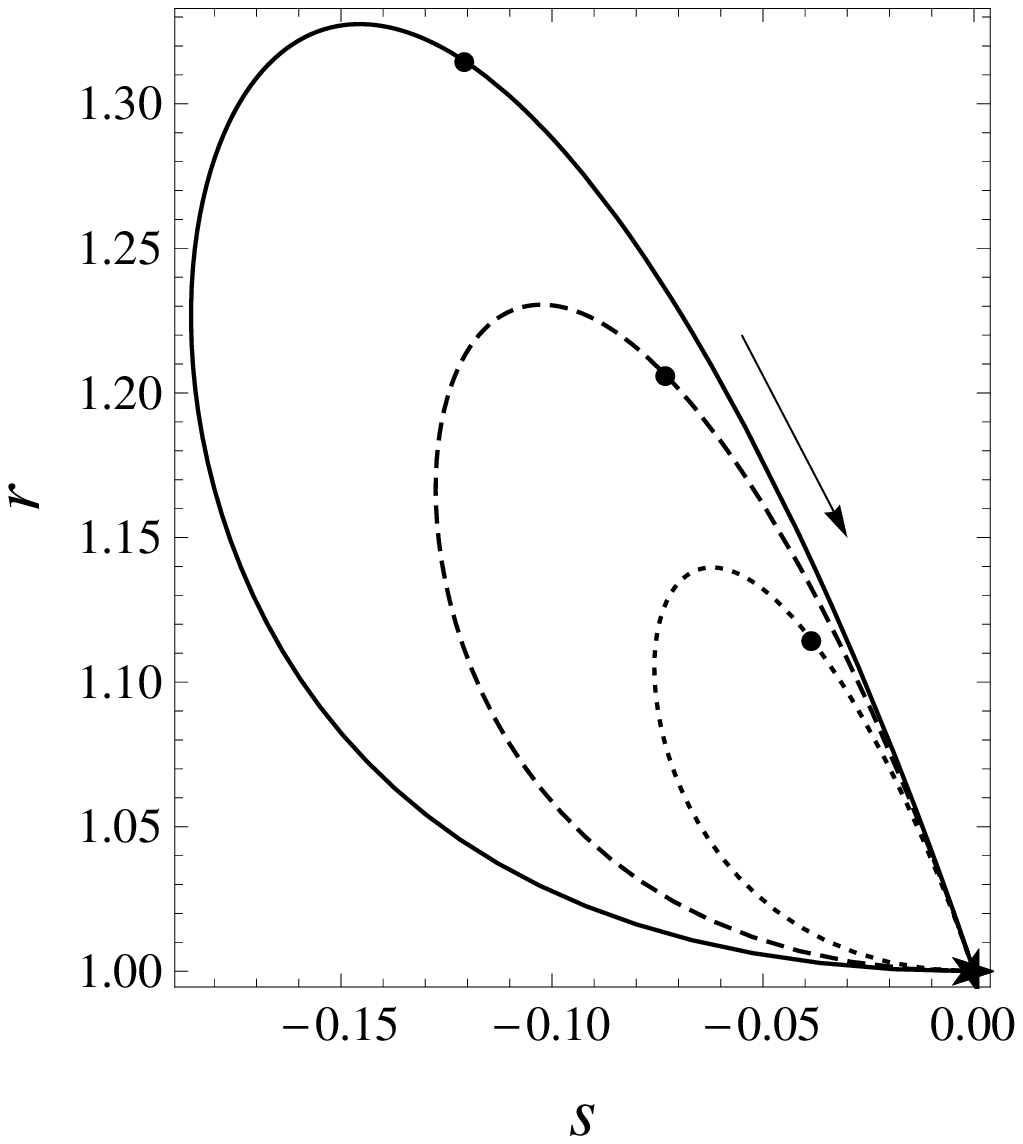}%
\hspace{.1\textwidth}%
\includegraphics[width=.45\textwidth]{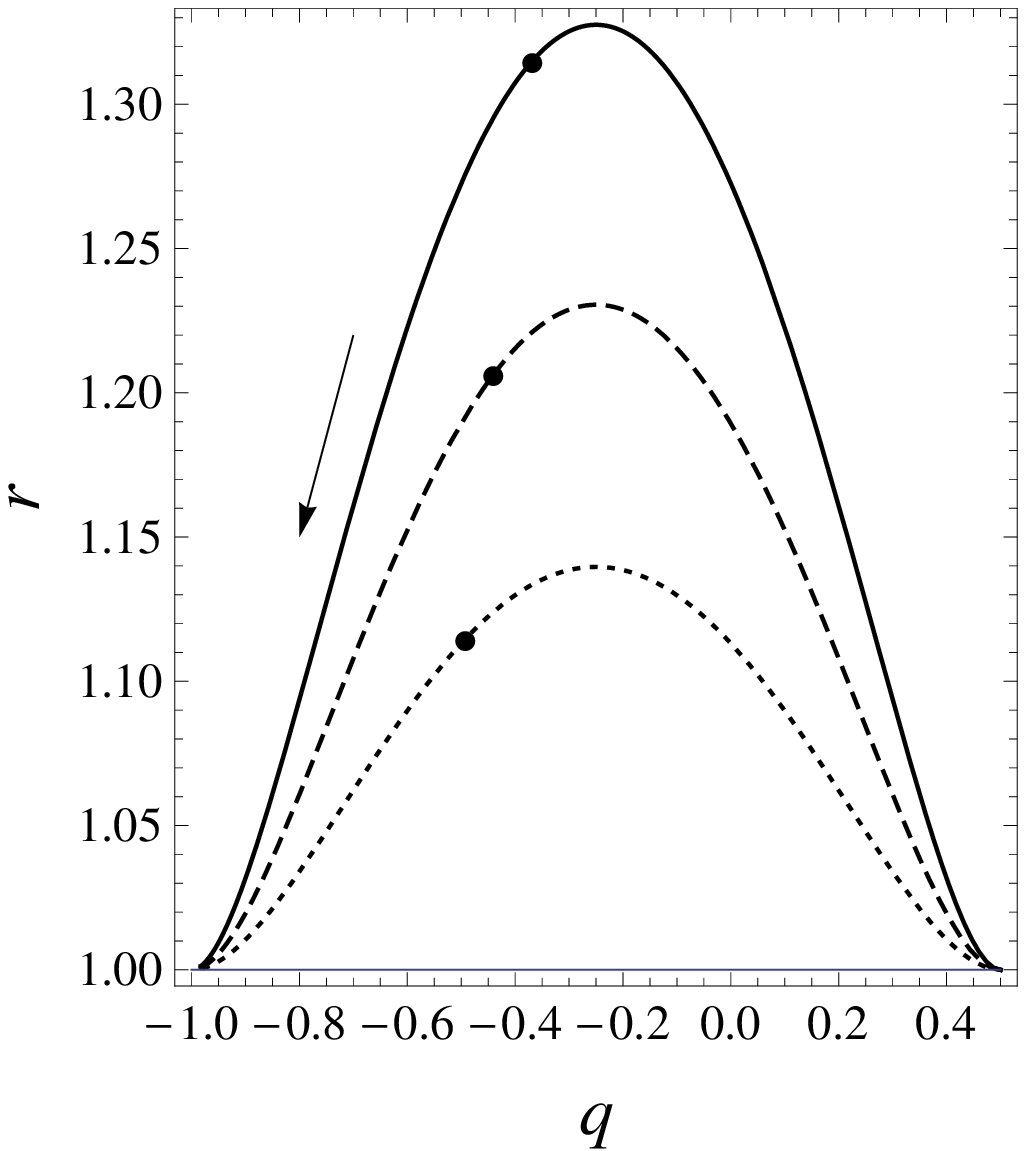}%
\caption{The statefinder evolution diagrams for the SCG model with
the dust-like normal component. The quantity $k$ varies as  0.653
(solid line), 0.488 (dashed line) and 0.345 (dotted line). Dots
mark the current values of the statefinder parameters and arrows
show the evolution direction of the statefinder trajectories. The
star denotes the $\Lambda$CDM location.}\label{FigDust}
\end{figure}

In the figure~\ref{FigDust} we plot the evolution trajectories in
the $s-r$ and $q-r$ planes assuming the current DE fraction
$\Omega_{\Lambda 0}=0.7$ and varying $k$ as 0.652, 0.488 and 0.345
that corresponds to $z_\text{T}=$0.3, 0.4 and 0.5 respectively.

The trajectory in the $s-r$ plane begins and ends  at the same
point corresponding to the $\Lambda$CDM. This is the feature of
the SCG model with the pressureless DM component. It does not take
into account the radiation  and therefore the total pressure is
the negative DE pressure and $q < 1/2$ through all the universe
evolution. The value $k=0$ corresponds to the transition redshift
of $\Lambda$CDM scenario
$z_{\text{T}}=(\kappa_0/2)^{-1/3}-1\approx 0.671$. In this case
the loop in the figure~\ref{FigDust} degenerates into the fixed
point \{0,1\} and the SCG model coincides with $\Lambda$CDM.

\begin{figure}[t]
\includegraphics[width=.45\textwidth]{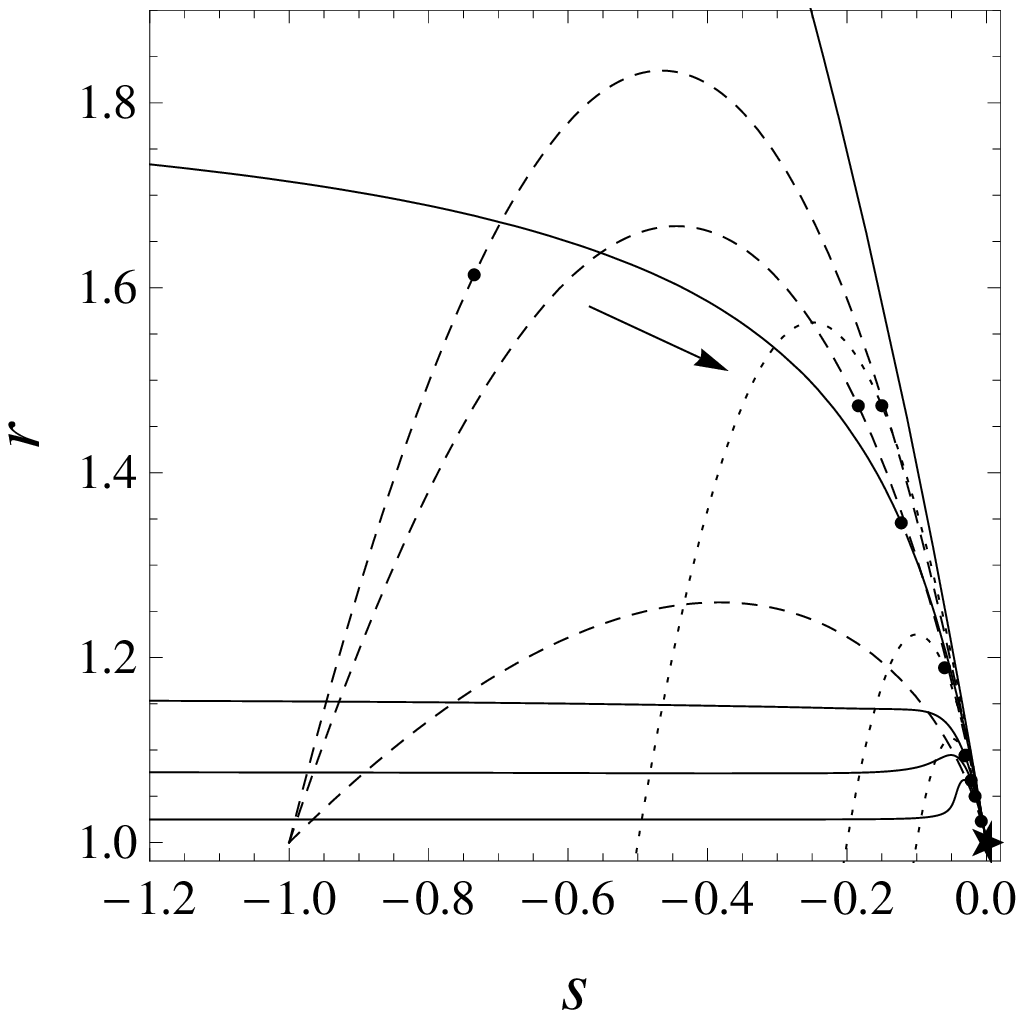}%
\hspace{.1\textwidth}%
\includegraphics[width=.45\textwidth]{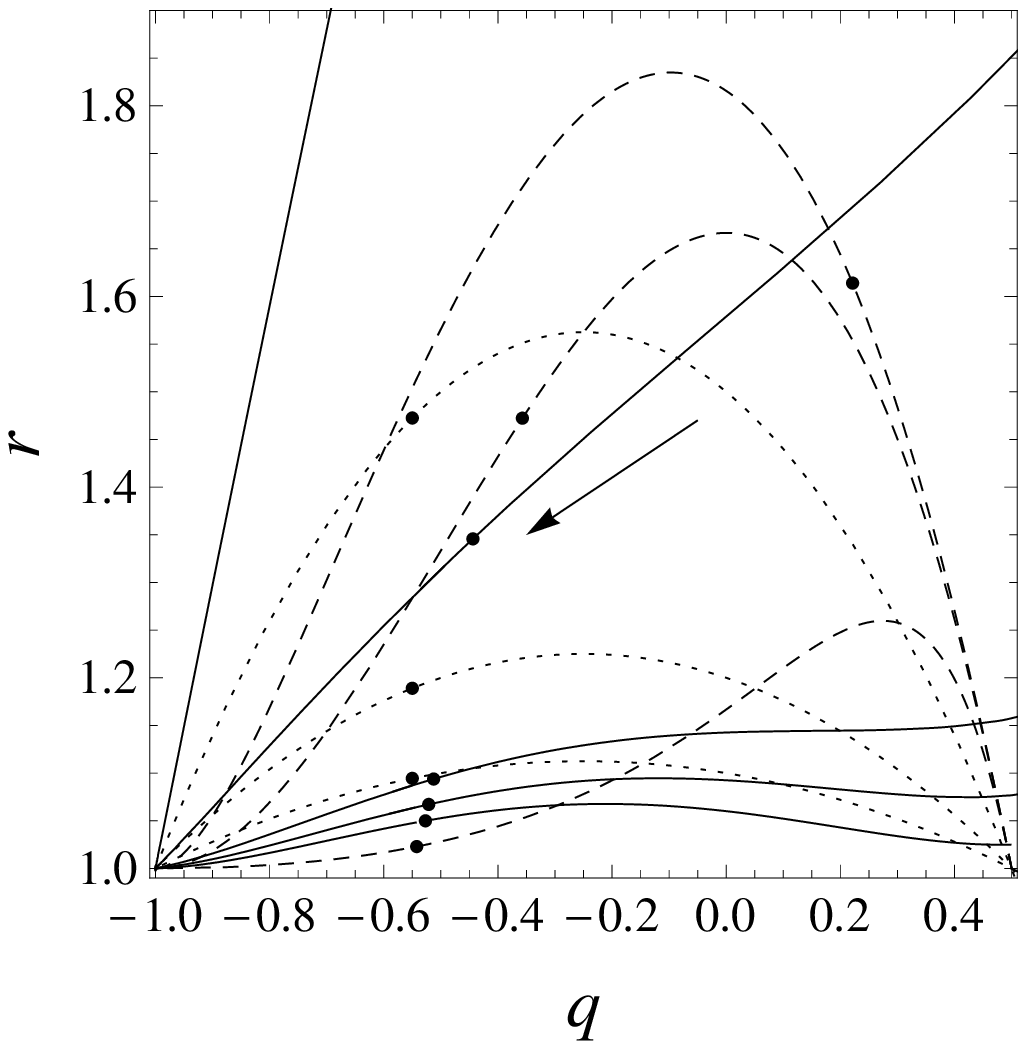}%
\caption{The statefinder evolution diagrams for the SCG model for
$k=0.2$ and the different values of $\polytrop=1,\ 5,\ 25,\ 50,\
150$ (the solid lines from top to down). The trajectories for the
model of the Chaplygin gas with CDM (the dashed lines corresponds
to $\kappa=0.5,\ 1,\ 5$ from top to down) and the GCG model (the
dotted lines corresponds to $\alpha=1,\ 0.5,\ 0.05$ from top to
down) is added for comparison. Dots mark the current values of the
statefinder parameters and arrows show the evolution direction of
the statefinder trajectories. The star denotes the $\Lambda$CDM
location.}\label{FigSoft}
\end{figure}

When $\polytrop$ is finite the pressure of DM is positive and the
expressions (\ref{rSCGgen}) and (\ref{sSCGgen}) are directly used
to calculate the statefinder evolution based on the numerical
solution of eqs.~(\ref{Eq1}) and (\ref{Eq2}). In the
figure~\ref{FigSoft} we plot the evolution trajectories $r(s)$ and
$r(q)$ for the various values of the quantity $\polytrop$.

At early times the DM  pressure and energy density exceed the DE
ones and ensure that the total pressure is positive. At the
present stage of DE dominance the universe expands with
acceleration driven by the negative total pressure. Between these
regimes there is a moment of time when the negative pressure of DE
is balanced by the positive pressure of DM. In this point the
total pressure is zero and $s\to\infty$. In fact, this point
exists in the universe evolution even though DM is pressureless
when we take into account the whole energy content of the
universe. Moreover the most considerable contribution in the
positive pressure at this stage is given by the radiation. The
non-zero DM pressure only shifts the moment of time when
$p_\text{tot}=0$. Trajectories in the figure~\ref{FigSoft} are
shown after this moment to focus the attention to the problem of
the recent accelerated expansion of the universe.

Another quantity in the SCG model, $k$, realizes the indirect
interaction between two components. It is clear from
eqs.~(\ref{Eq1}) and (\ref{Eq2}) that varying $k$ can be
counterbalanced by rescaling of the scale factor $a$ leaving the
equations invariant. However, for a fixed ratio between the DM and
DE energy densities different $k$ are corresponded to different
trajectories. Although the equations can be solved for any $k$, it
is restricted by the observational estimations of the transition
redshift $z_\text{T}$ \cite{Wang1}. For the current DE content
$\Omega_{\Lambda 0}=0.7$ the value of $k$ does not exceed 0.652
determined by the limiting case of the pressureless normal
component. The figure~\ref{FigSoft} depicts the evolution curves for $k=0.2$ and varying $k$ gives the similar plots which are different only in a quantitative sense.

The figure~\ref{FigSoft} also contains the evolution trajectories
for two alternative models with the Chaplygin gas to study
differences in their statefinder evolution. The former describes
the universe with DE obeying the Chaplygin equation of state
$p_\Lambda=-A/\rho_\Lambda\,$,
and CDM. This is two-component model without interaction between
their parts, where the energy densities of the Chaplygin gas and
CDM evolve according to
\beq\label{KamenRho}
\rho_\Lambda=\sqrt{A+Ba^{-6}}\, , \qquad \rho_\text{m}=Ca^{-3}\, .
\eeq
The statefinder diagnostic of this model was carried out in
\cite{Kam-sf-in-Chap,Alam-sf}. Substituting (\ref{KamenRho}) into
(\ref{qPress})--(\ref{sPress}) one can obtain the explicit
dependence
\beq
r(s)=1-\frac{9}{2}\,\frac{s(s+1)}{1+\kappa\sqrt{-s}}\, ,\qquad%
q(s)=\frac{1}{2}\left( 1- \frac{3(s+1)}{1+\kappa\sqrt{-s}}
\right),
\eeq
where $\kappa=C/\sqrt B$ is the ratio between CDM and the
Chaplygin gas energy densities at the beginning of the
cosmological evolution.

The latter model is the generalized Chaplygin gas (GCG) with the
equation of state $p=-A/\rho^\alpha\, ,\ (0\leq\alpha\leq 1)$, and
the energy density evolves according to
\beq
\rho=\left( A+Ba^{-3(1+\alpha)} \right)^{1/(1+\alpha)} .
\eeq
It is also suggested that the energy density $\rho$ consists of
both vacuum and matter contributions. This is favorable to use the
GCG model for a DE and DM unification \cite{Bento&Bilic,Bento2}.

Statefinder parameters has explicit dependence
\beq
r(s)=1-\frac{9}{2}\,\frac{s(s+\alpha)}{\alpha}\, ,\qquad%
q(s)=\frac{1}{2}\left( 1- \frac{3(s+\alpha)}{\alpha} \right).
\eeq

It is obvious from the figure~\ref{FigSoft} that the evolution
trajectories are distinct for all three models and they converge
into the same point at the far future. Note that the models become
hard to be distinguished at the late stage in the $s-r$ diagram
while they remain quite different in the $q-r$ plane.

\begin{figure}[t]
\includegraphics[width=.43\textwidth]{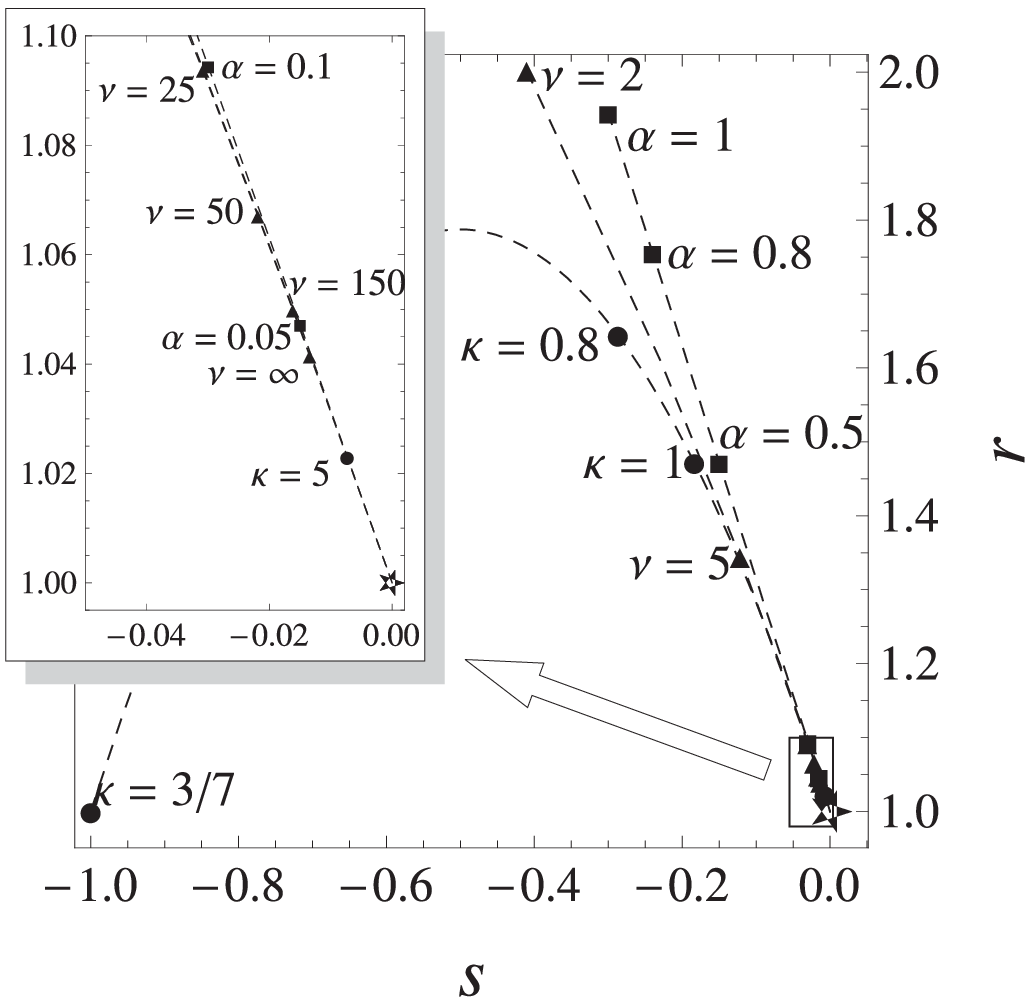}%
\hspace{.1\textwidth}%
\includegraphics[width=.47\textwidth]{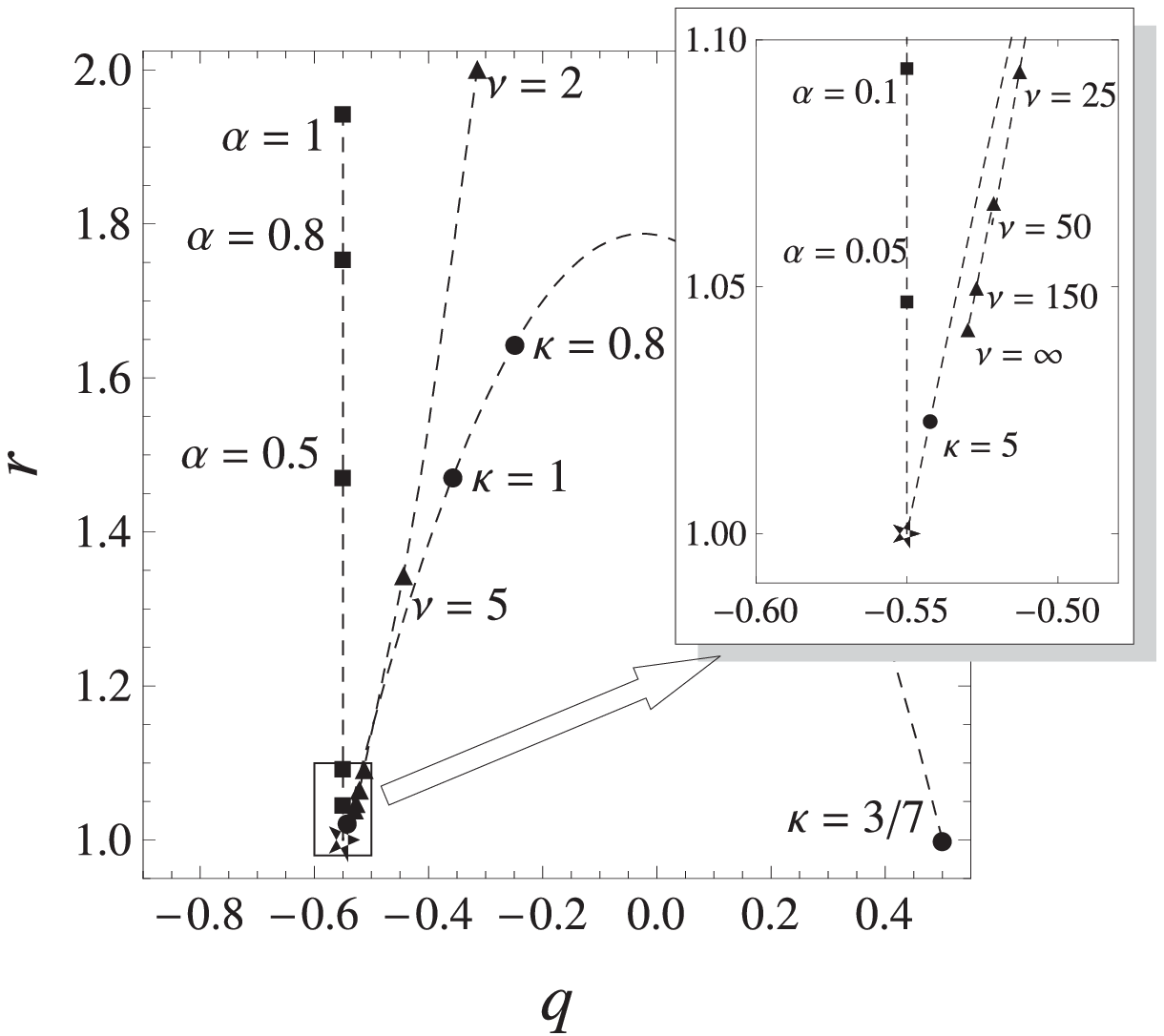}%
\caption{The current statefinder locations. The diagrams show the
current positions for the Chaplygin gas model with CDM (circles),
the GCG model (squares) and the SCG model (triangles). The star
denotes the $\Lambda$CDM location. The areas bordered by the
rectangles are enlarged to resolve the values near the
$\Lambda$CDM location (star).}\label{FigChaplygin}
\end{figure}

Nevertheless these trajectories are not of determining
significance in themselves. The current statefinder values are of
primary importance to differentiate cosmological scenarios from
$\Lambda$CDM and to impose restrictions on the models. The
figure~\ref{FigChaplygin} shows the current statefinder locations
for the models with the Chaplygin gas assuming that the current DE
density $\Omega_{\Lambda 0}=0.7$.

It is the sole parameter $\kappa$ determining the statefinder
evolution trajectory and fixing  the modern statefinder values in
the model with the pure Chaplygin gas. We start with the value of
$\kappa=3/7$ which coincides with the current ratio between DM and
DE energy densities  that leads to the modern statefinder values
located far from the $\Lambda$CDM point. It is easy to see that
already for $\kappa=5$ the modern location is fairly close to
$\Lambda$CDM. In the GCG model one uses the decomposition of the
energy density into two components
\beq\label{splitGCG}
\rho=\rho_\Lambda+\rho_\text{m}\quad\text{and}\quad%
p= -\rho_\Lambda
\eeq
proposed in \cite{Bento2} to fix the current statefinder position.
The figure~\ref{FigChaplygin}  shows that it approaches the
$\Lambda$CDM point as $\alpha$ tends to zero.

The current values for the SCG model are given for fixed $k=0.2$
and different $\polytrop$. As one would expect, they are far from the
$\Lambda$CDM point as well as from other models concerned when
$\polytrop$ is small, and tend to the $\Lambda$CDM fixed point when
$\polytrop$ increases.

The $s-r$ diagram demonstrates the very close trajectories for all
three models near the $\Lambda$CDM point. In contrast,  the $q-r$
diagram allows to differentiate between these models. In the GCG
model the current deceleration factor is the same as in
$\Lambda$CDM for any $\alpha$ owing to the selected decomposition
(\ref{splitGCG}). It implies that today's statefinder locations
for different $\alpha$ lies along the vertical line $q=\left( q_0
\right)_{\Lambda\text{CDM}}=0.55$ in the $q-r$ diagram.

In the model with the pure Chaplygin gas
\beq
q_0=\left( q_0 \right)_{\Lambda\text{CDM}} + \frac{3}{2}\,
\frac{\kappa_0^2/\kappa^2}{1+\kappa_0}
\eeq
and it is to be found resting on the parabola to the right of the
vertical line.

When $\polytrop$ increases the current statefinder location line in the
SCG model come close to the parabola since large $\polytrop$ means that
the normal component behaves like CDM. Decreasing $k$ implies
attenuation of interrelation between the components of SCG that
also leads to the approach of the SCG and pure Chaplygin gas
models and further degeneration SCG into $\Lambda$CDM.

\begin{figure}[t]
\includegraphics[width=.45\textwidth]{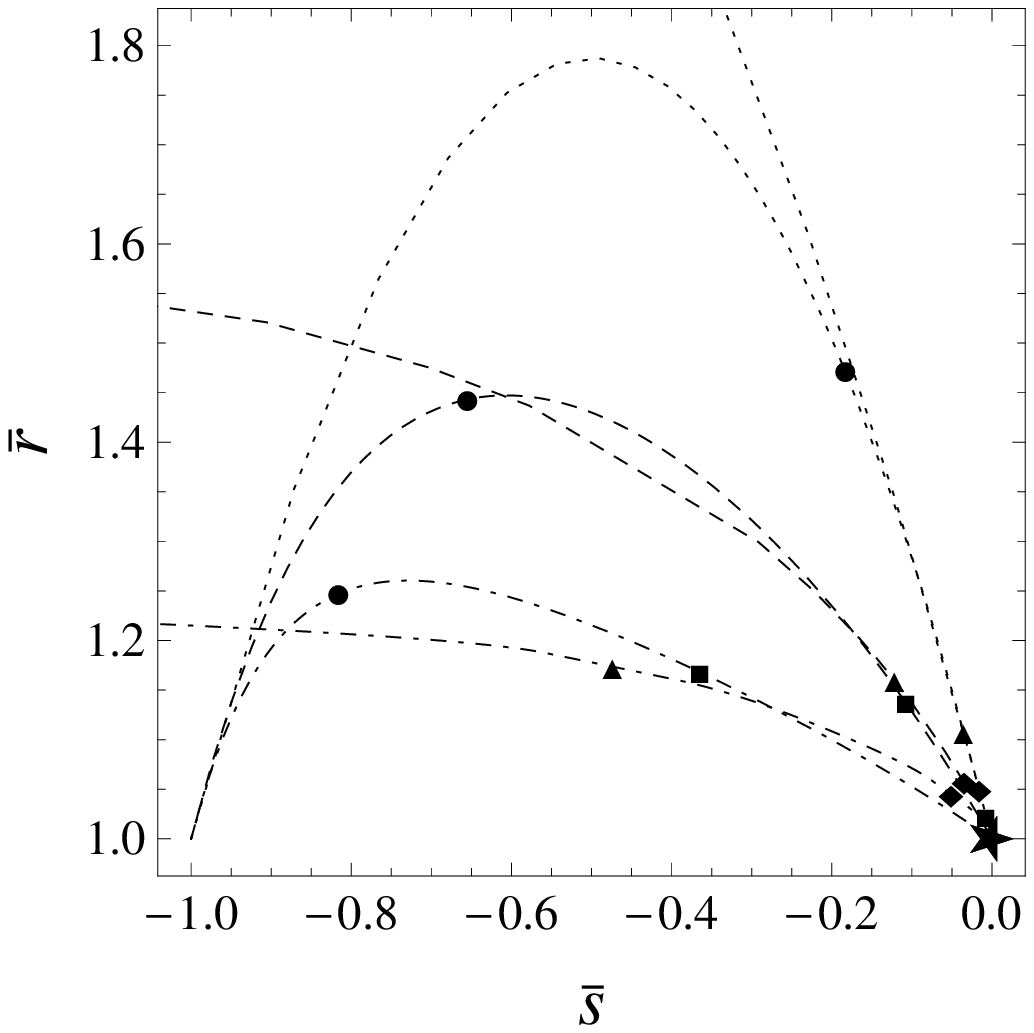}%
\hspace{.1\textwidth}%
\includegraphics[width=.45\textwidth]{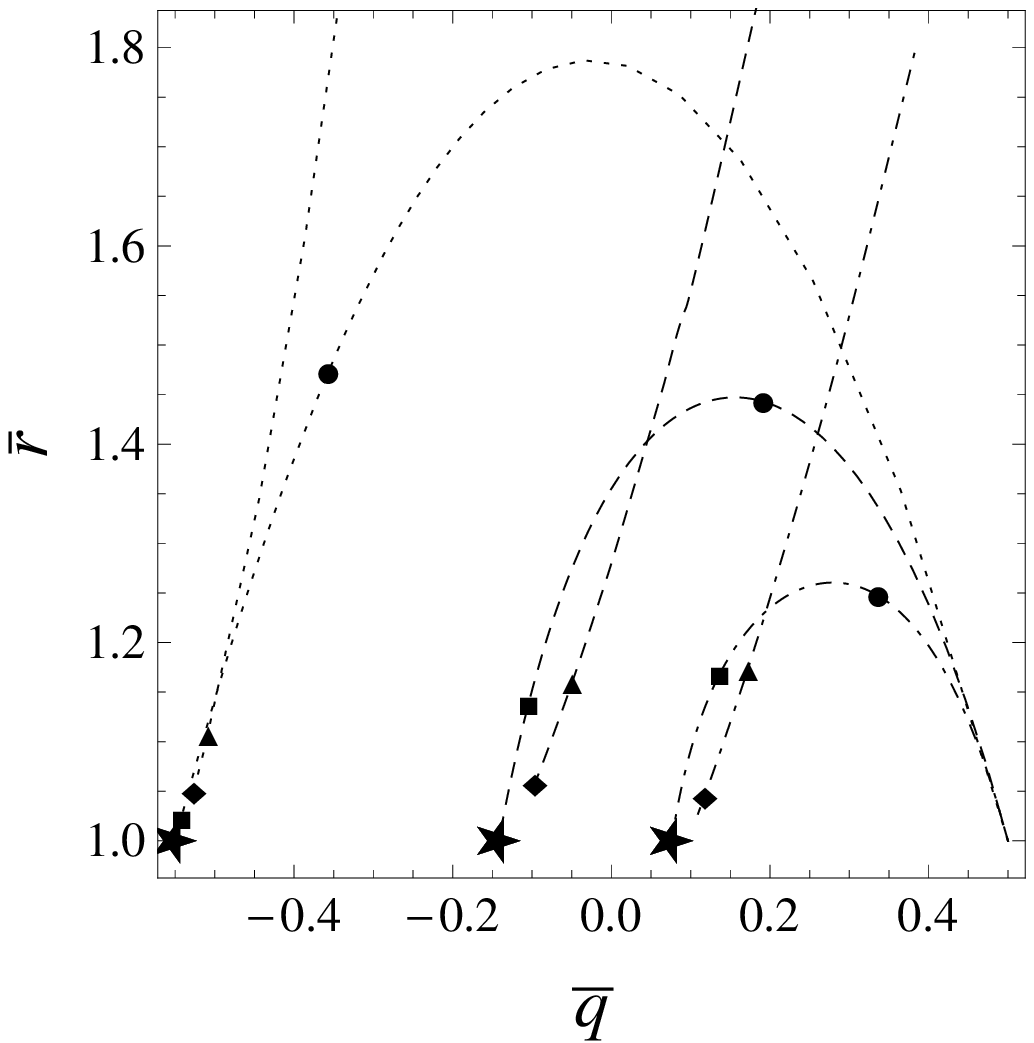}%
\caption{The integrated statefinder locations. The diagrams show
the integrated statefinder indices defined by (\ref{sfInegr}) for
the Chaplygin gas model with CDM (parabolas) for different
$\kappa$ and the SCG model for different $\polytrop$ at $k=0.2$.
The dashed lines corresponds to $z_\text{max}=1$, the dot-dashed
lines corresponds to $z_\text{max}=2$, and the dotted lines
contain the current locations. The values for $\kappa=1$ (circles)
and 5 (squares), and $\polytrop=20$ (triangles) and 150
(rhombuses) are marked out for detail comparison. The stars denote
the $\Lambda$CDM locations.}\label{FigSFbar}
\end{figure}

\begin{figure}[t]
\includegraphics[width=.45\textwidth]{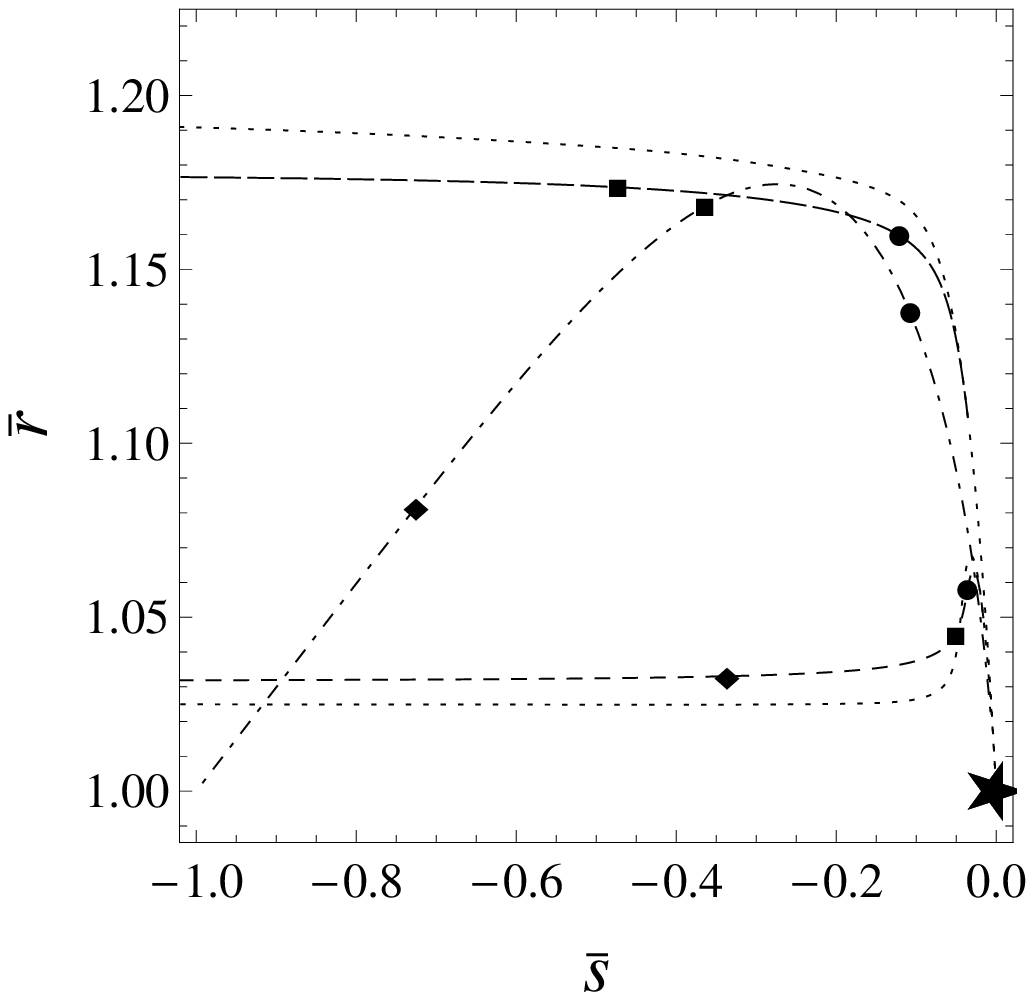}%
\hspace{.1\textwidth}%
\includegraphics[width=.45\textwidth]{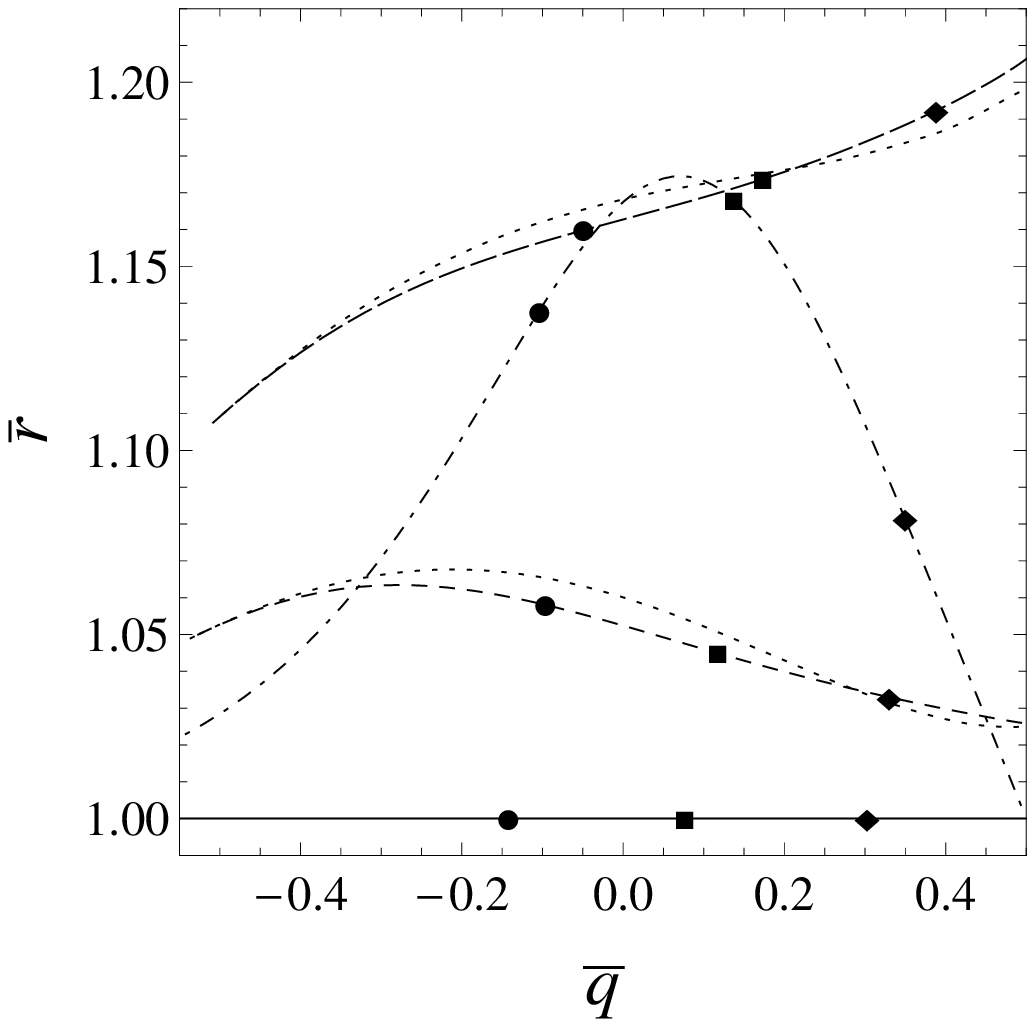}%
\caption{$z_\text{max}$-dependence of the integrated statefinder quantities. The dot-dashed lines correspond to the Chaplygin gas model with CDM for $\kappa=5$. The dashed lines correspond to SCG model for $\polytrop=20$ (long dashes) and 150 (short dashes), and the dotted lines represent the statefinder evolution trajectories for the same parameters. $\Lambda$CDM is shown as the star in the left panel and the horizontal solid line in the right one. The values for $z_\text{max}=1$ (circles), 2 (squares) and 5 (rhombuses) are marked out for detail comparison.
}\label{FigSFzmax}
\end{figure}

In order to distinguish these models with confidence we use the
following integrated quantities
\beq\label{sfInegr}
\bar q = \frac{1}{z_\text{max}} \int\limits_0^{z_\text{max}}\!\! q
\d z\, ,
\quad %
\bar r = \frac{1}{z_\text{max}} \int\limits_0^{z_\text{max}}\!\! r
\d z\, ,
\quad %
\bar s = \frac{1}{z_\text{max}} \int\limits_0^{z_\text{max}}\!\! s
\d z\, ,
\eeq
introduced in \cite{Alam-sf} to take into account a previous DE
evolution. The figure~\ref{FigSFbar} depicts the lines passing
through the points associated with the pairs $\{ \bar s , \bar r
\}$ and $\{ \bar q , \bar r \}$ for different values of $\kappa$
in the pure Chaplygin gas model and for different $\polytrop$ in
the SCG model. The current location lines are added for
comparison. It is apparent that the certain values of $z_\text{max}$
considerably separates the models even though their current
statefinder positions are almost indistinguishable. It is clear that in general an opposite situation can also take place. Because of this ambiguity the trend of $z_\text{max}$-dependence should be specially considered. The figure~\ref{FigSFzmax} shows the $\bar s - \bar r$ and $\bar s - \bar r$ diagrams governing by $z_\text{max}$. Magnitudes of the quantities (\ref{sfInegr}) are less then the maximal corresponding statefinder values in the range $[0,z_\text{max}]$, and the evolution curves in the figure~\ref{FigSoft} are quite smooth, therefore the integrated statefinder trajectories are similar to the evolution ones and their efficiency could be developed in different ways. Since the SCG trajectories in the $\bar q - \bar r$ plane for large $\polytrop$ become closer to the $\Lambda$CDM line at greater $z_\text{max}$ then to increase $z_\text{max}$ is not so reasonable as to improve a statistical accuracy through a larger number of SNIa in the observational redshift range. To the contrary, the difference between the SCG model and  $\Lambda$CDM in the $\bar s - \bar r$ diagram is enhanced when $z_\text{max}$ increases. It is primarily caused by a growing magnitude of the parameter $s$ in the SCG model up to the instant when $p_\text{tot}=0$ while $\Lambda$CDM is represented as the fixed point $\{0,1\}$ as in the current statefinder diagram. This advantage is favorable to distinguish the competing models with more confidence using observations of SNIa at higher redshifts. Similar estimations were carried out in \cite{Alam-sf}, where the authors revealed that the discriminatory ability of the statefinders varies with redshift and showed that it improves when $q$, $r$ and $s$ in (\ref{sfInegr}) are integrated over different redshift ranges.

\section{Conclusion}

In this letter the SCG model is studied from the statefinder
viewpoint. This model describes the dark sector of the universe as
a matter that behaves as DE while it is in the ground state and as
DM when it is in the excited state. Cosmological dynamics is
described in the framework of the relativistic superfluid model
therefore the interaction between DE and DM is implicitly involved
into the conservation laws (\ref{EnergyConservation}) and
(\ref{ParticleConservation}).

The condensate possesses the equation of state of the Chaplygin
gas but the universe evolution provided by this matter is
different from the two-component model with the Chaplygin gas and
CDM as well as from the GCG model used for unifying DE and DM. The
discrimination is obviously demonstrated in the statefinder
evolution diagrams. The diagrams show that for fixed ratio between
DM and DE energy densities two quantities determine the trajectory
and the current statefinder location. The former, $\polytrop$,
governing the DM equation of state ought to be quite large to
correspond to the cosmological observations. It implies that the
pressure of the normal component is small and it behaves like CDM.
From the superfluid standpoint it means that the second sound speed
is small too and this inference should be taken into account for any
(realistic or simulative) DM equation of state.
The latter, $k$, interrelating the DM and DE is restricted for the
universe commenced to accelerate before now. The limiting case of
infinite $\polytrop$ and $k=0$ corresponds to $\Lambda$CDM and
establishes the maximal value for the transition redshift
$z_\text{T}=(z_\text{T})_{\Lambda\text{CDM}}=0.671$ in the SCG
model.

Near the $\Lambda$CDM fixed point the SCG and pure Chaplygin gas
models are close together but they can be separated if the earlier
evolution is taken into account.
It is found that the better evaluation could be developed at lower redshifts for the parameters $\bar q$ and $\bar r$, and at higher redshifts for $\bar s$. The same inference was made in \cite{Alam-sf} on the statefinder analysis of a number of models.
As it is shown in \cite{Alam-sf}
the observational data from the SNAP type experiments are in
reasonably good agreement  with $\Lambda$CDM and rule out the
models whose current statefinder values locate far from the
$\Lambda$CDM point. This is the reason that an effect of the
previous DE evolution takes on great importance. It is hoped that
the future high-z supernova observations will provide new data to
clarify the essence of DE.

\end{document}